\documentclass[12pt]{article}
\usepackage{graphicx}
\usepackage{palatino}
\usepackage{color}
\usepackage{amsmath}
\usepackage{amsfonts}
\usepackage{amssymb}
\usepackage{dsfont,mathrsfs}
\usepackage{geometry}
\usepackage{setspace}
\usepackage{cancel}
\usepackage{bm}
\usepackage{epstopdf}
\usepackage{subfig}

\def\bea {\begin{eqnarray}}
\def\eea {\end{eqnarray}}

\def\be {\begin{equation}}
\def\ee {\end{equation}}

\definecolor{checkcolor}{rgb}{0.75, 0.75, 0.75}
\newsavebox{\definitionbox}
{\end{minipage}\end{lrbox}%
\begin{center}{\colorbox{checkcolor}{\usebox{\definitionbox}}}%
\end{center}}

\title{Charge correlations using balance functions of identified particles in Pb–Pb collisions at $\sqrt{s_{NN}}$ = 2.76 TeV with ALICE}

\author{Sk Noor Alam \\
        for the ALICE Collaboration \\
         sk.noor.alam@cern.ch\\
        Variable Energy Cyclotron Centre, HBNI,Kolkata India}

\date{\today}%

\begin{document}

\maketitle
\sloppy

 \begin{abstract}
By studying the balance functions of several hadronic species, one can gain insight into the chem-ical evolution of the Quark–Gluon Plasma and radial flow. In a picture of early hadronisation,pairs of particles and anti-particles (created at the same space–time point) are separated furtherin rapidity due to the higher initial temperature and diffusive interactions with other particles.Therefore delayed hadronisation will lead to strong correlations in rapidity in the final state.  Inthis work, balance functions for identified charged-pion pairs measured by the ALICE detector atthe Large Hadron Collider (LHC) in Pb–Pb collisions at $\sqrt{s_{NN}}$ = 2.76 TeV are presented. These balance functions are presented in relative rapidity ($\Delta y$) and relative azimuthal angle ($\Delta \varphi$).  It isobserved that the charged-pion balance function widths in $\Delta y$ and $\Delta \varphi$ get narrower in central Pb-Pb collisions compared to peripheral collisions.  The findings in this analysis are consistent withthe effects of delayed hadronisation and radial flow.

\end{abstract}


\section{Introduction}
The formation of a new state of matter produced in relativistic heavy-ion collisions called the Quark-Gluon Plasma (QGP) has been hypothesised to exist during the early stages of these colli-sions. The balance function (BF) is sensitive to the effects of electric charge conservation among the produced particles in high energy collisions \cite{Ref:1st,Ref:2nd,Ref:3rd}.  When measuring the BF as a function of relative rapidity/pseudorapidity (y, $\eta$) and azimuthal angle ($\varphi$), it is shown that a late stage hadronisation is characterised by tightly correlated charge-anticharge pairs \cite{Ref:4th}.  Early stage hadronisationis expected to result in a broad BF, while late stage hadronisation leads to a narrower distribution.The BF is constructed by the normalised net correlation between oppositely charged particles andis defined as \cite{Ref:4th, Ref:5th}

\begin{equation}
B = \frac{1}{2} \Bigg [ \frac{\langle N_{(a,b)} \rangle - \langle N_{(a,a)} \rangle }{\langle N_a \rangle}  + \frac{\langle N_{(b,a)} \rangle - \langle N_{(b,b)} \rangle }{\langle N_b \rangle}  \Bigg]
\label{Eq:BFEq}
\end{equation}

where a and b could be different kinds of particles with positive and negative charges. For example, a could refer to all negative particles and b to all positive particles. Here, $N_{a,b}$ counts pairs of opposite charges satisfying a criterion that their relative rapidity/pseudorapidity or $\varphi$ is within a specified range, whereas $N_a$, $N_b$ are the number of positive or negative particles in the chosen phase space. Here the angular brackets represent averaging over the events. The BF is measured by combining two types of particles called trigger and associated particles with transverse momenta $\textit{p}_{T,trig}$ and $\textit{p}_{T,assoc}$ . \\
The single terms were constructed in two different ways, the associated yield per trigger particle ~\cite{Ref:6th, Ref:7th} and the number correlation function $R_2$ ~\cite{Ref:8th}. In this analysis a pseudorapidity range ($\eta$) between -0.8 and 0.8 and $\textit{p}_{T,trig}$ and $\textit{p}_{T,assoc}$ range between 0.2 and 1.4 GeV/c was used. First, a two-dimensional correlation function is generated in ($\Delta \eta , \Delta \varphi$) which is then projected on $\Delta \varphi$ and $\Delta \eta$ for further studies. The BF correlation follows a characteristic structure with a peak at the near side ($\Delta \eta$ = 0 , $\Delta \varphi$ = 0). By studying BFs involving different hadronic species, and calculating the width of the BF, one can obtain insight the chemical evolution of the QGP. The BF widths are also sensitive to amount of radial flow in heavy-ion collisions ~\cite{Ref:1st}.

\section{Dataset, event and track selection}
The extraction of the BF with identified particles was performed on the Pb$\textendash$Pb dataset at
$\sqrt{s_{NN}}$ = 2.76 TeV, which was taken in 2010. Approximately 12 M events were used. Only events with a reconstructed vertex and with z position of the vertex $\vert V_{z} \vert < $  10 cm were analysed. The events were divided in centrality classes, spanning $0 \textendash 80\%$ of the total inelastic cross section. The most central events were analyzed in 5$\%$ bins ($0 \textendash 5\%$ and $5 \textendash 10 \%$) and the more peripheral in 10$\%$ bins. The centrality of an event was estimated by using the multiplicity distribution of signals from the V0 detectors. Primary particle tracks are reconstructed, selected ~\cite{Ref:7th} and identified ~\cite{Ref:9th} using the Time Projection Chamber (TPC) and Time-Of-Flight detector (TOF). Here, the $n_{\sigma}$ method is used to identify particles with $n_{\sigma} < $ 3.
\begin{equation}
n_\sigma = \frac{Signal_{PID} - Signal_{Expected}}{\sigma}
\label{Eq:sigma}
\end{equation}

\begin{figure}[htp]
  \centering
  \begin{tabular}{cc}
    \includegraphics[width=80mm]{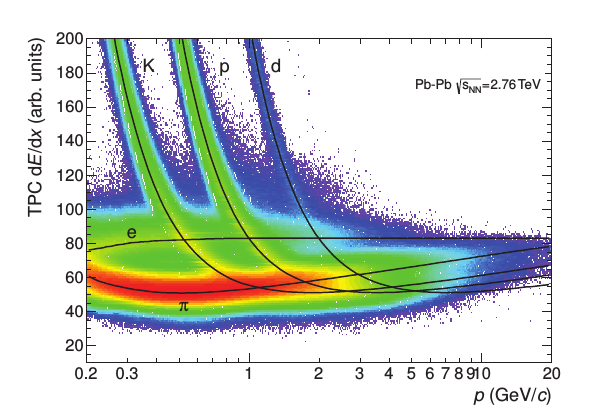}
    \includegraphics[width=80mm]{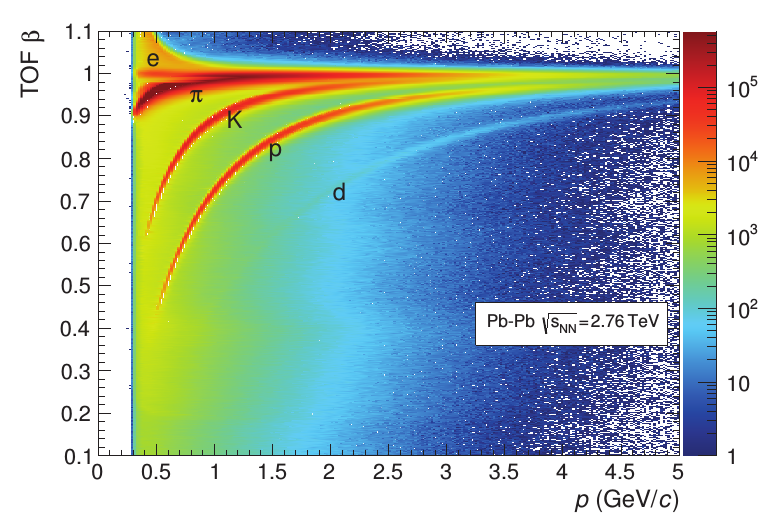} \\
  \end{tabular}
  \caption{Left: Specific energy loss (dE/dx) in the TPC vs. particle momentum in Pb-Pb
collisions at $\sqrt{s_{NN}}$ = 2.76 TeV. The lines show the parametrisation of the expected mean energy loss. Right: Distribution of $\beta$ as measured by the TOF detector as a function of momentum for particles reaching the TOF in Pb-Pb interactions. Figures are taken from \cite{Ref:9th}.}
  \label{fig:2}

\end{figure}

Here, signal means $\beta = \frac{v}{c}$ and relative energy loss w.r.t distance for TOF and TPC respectively. The TPC is used for identification of particles with $p_{T}$ of 0.2 to 0.6 GeV/c. The combination of TPC and TOF information have been used for identification of particles with $p_{T}$ of 0.6 to 1.4 GeV/c. Fig.\ref{fig:2} shows TPC and TOF signals extracted from Pb-Pb collisions with $\sqrt{s_{NN}}$ = 2.76 TeV energy. In this work, we have selected pions as trigger $\&$ associated particles. The correlation functions have been corrected for efficiency $\&$ purity obtained from simulated data.

\clearpage
\section{Results}
\begin{figure}[htp]
  \centering
  \begin{tabular}{cc}
  \includegraphics[width=65mm]{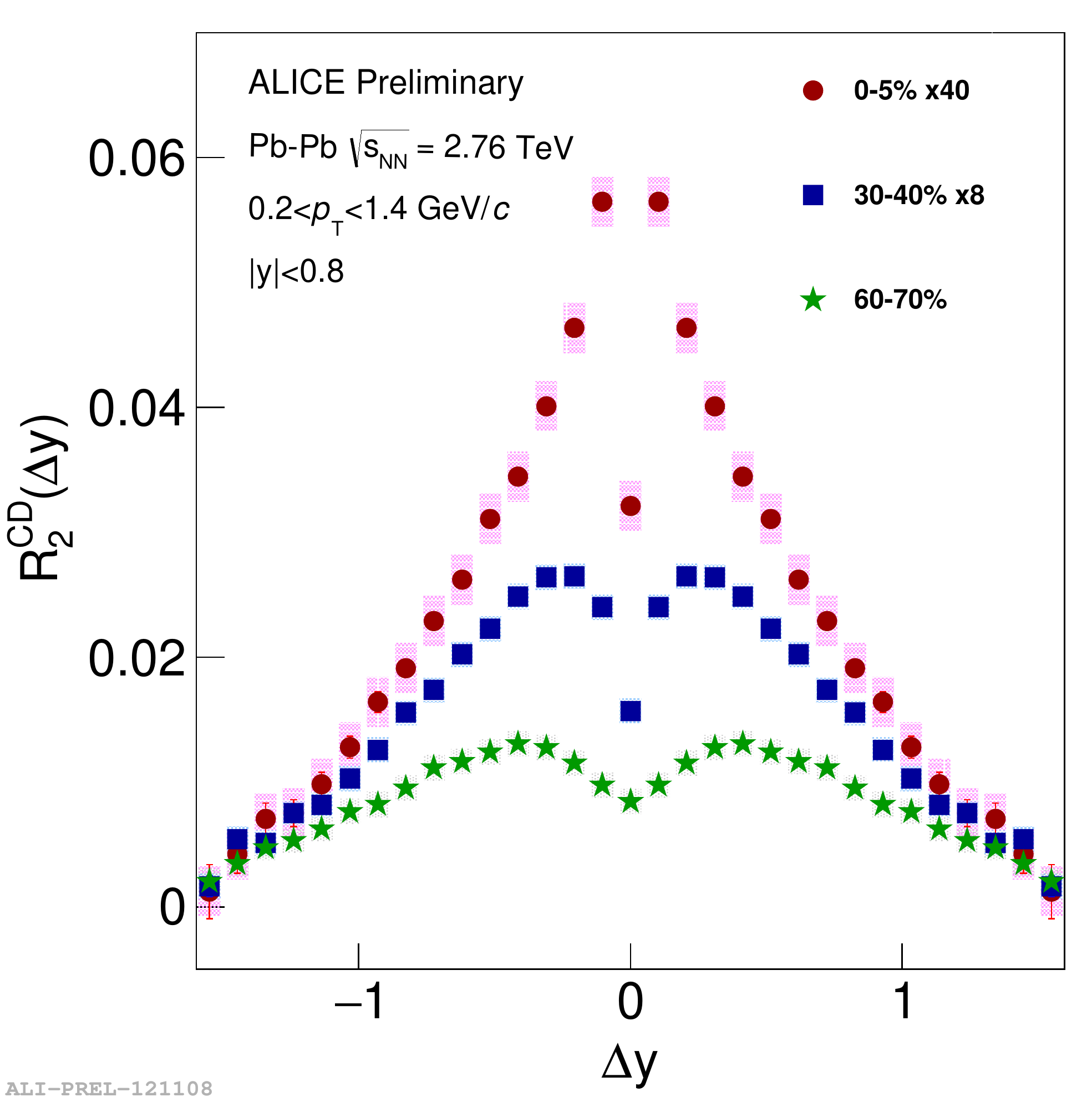}
    \includegraphics[width=65mm]{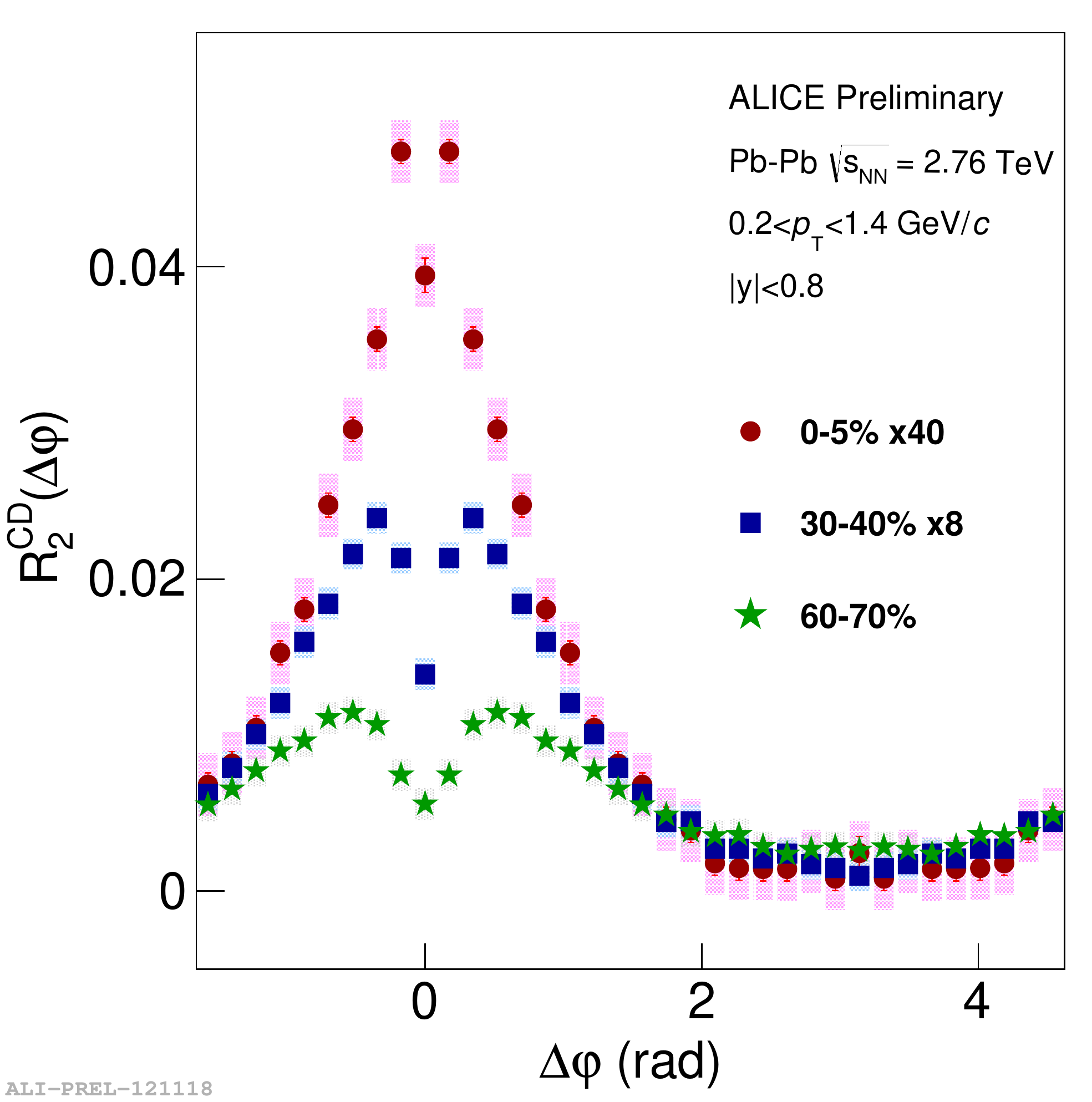} \\
  \end{tabular}
  \caption{The balance function versus y (left plot) and azimuthal angle ($\varphi$)   (right plot) for identified pion pairs from central and peripheral Pb-Pb collisions at $\sqrt{s_{NN}}$= 2.76 TeV}
  \label{fig:3}

\end{figure}

\begin{figure}[htp]
  \centering
  \begin{tabular}{cc}
  \includegraphics[width=65mm]{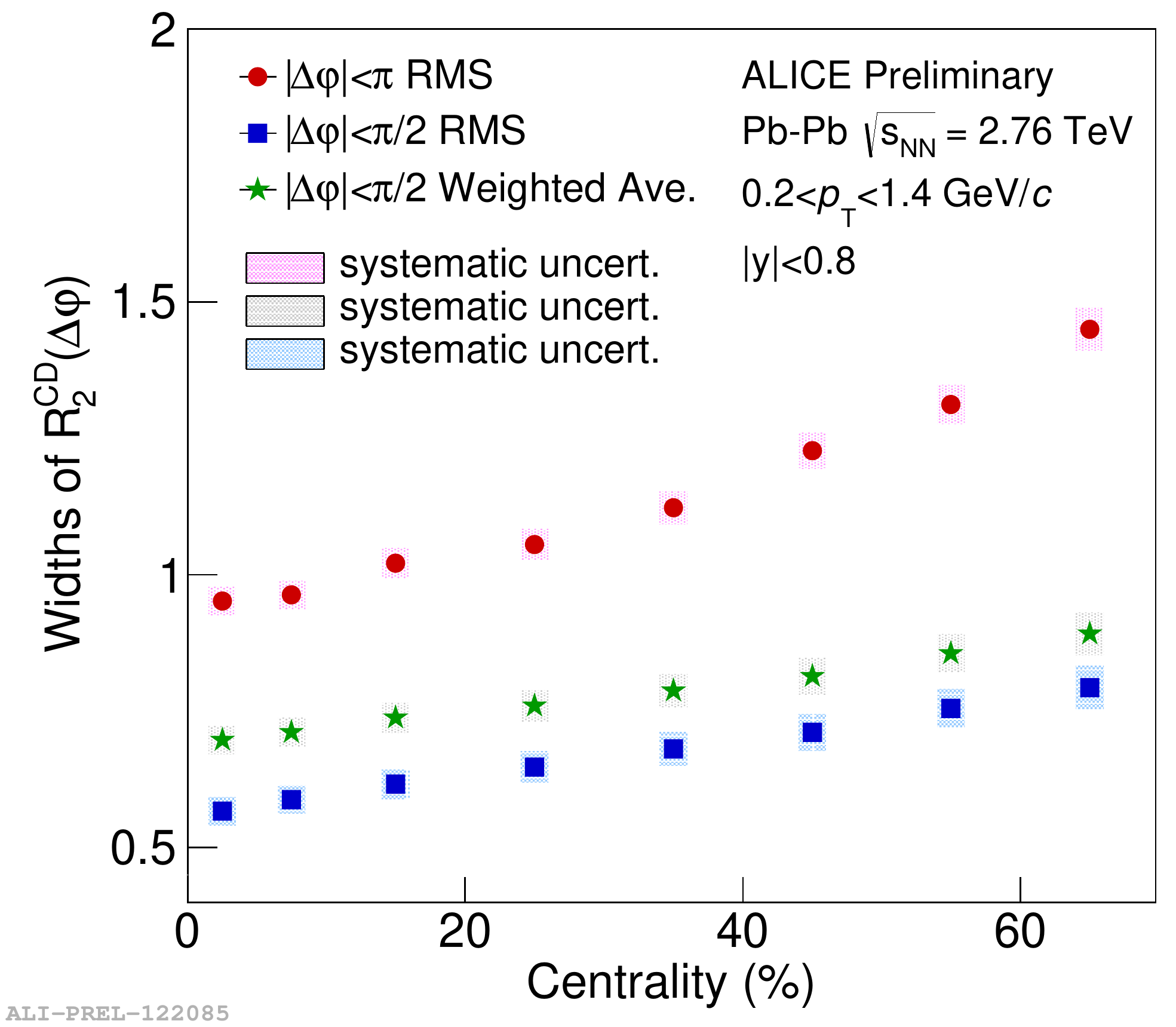}
    \includegraphics[width=65mm]{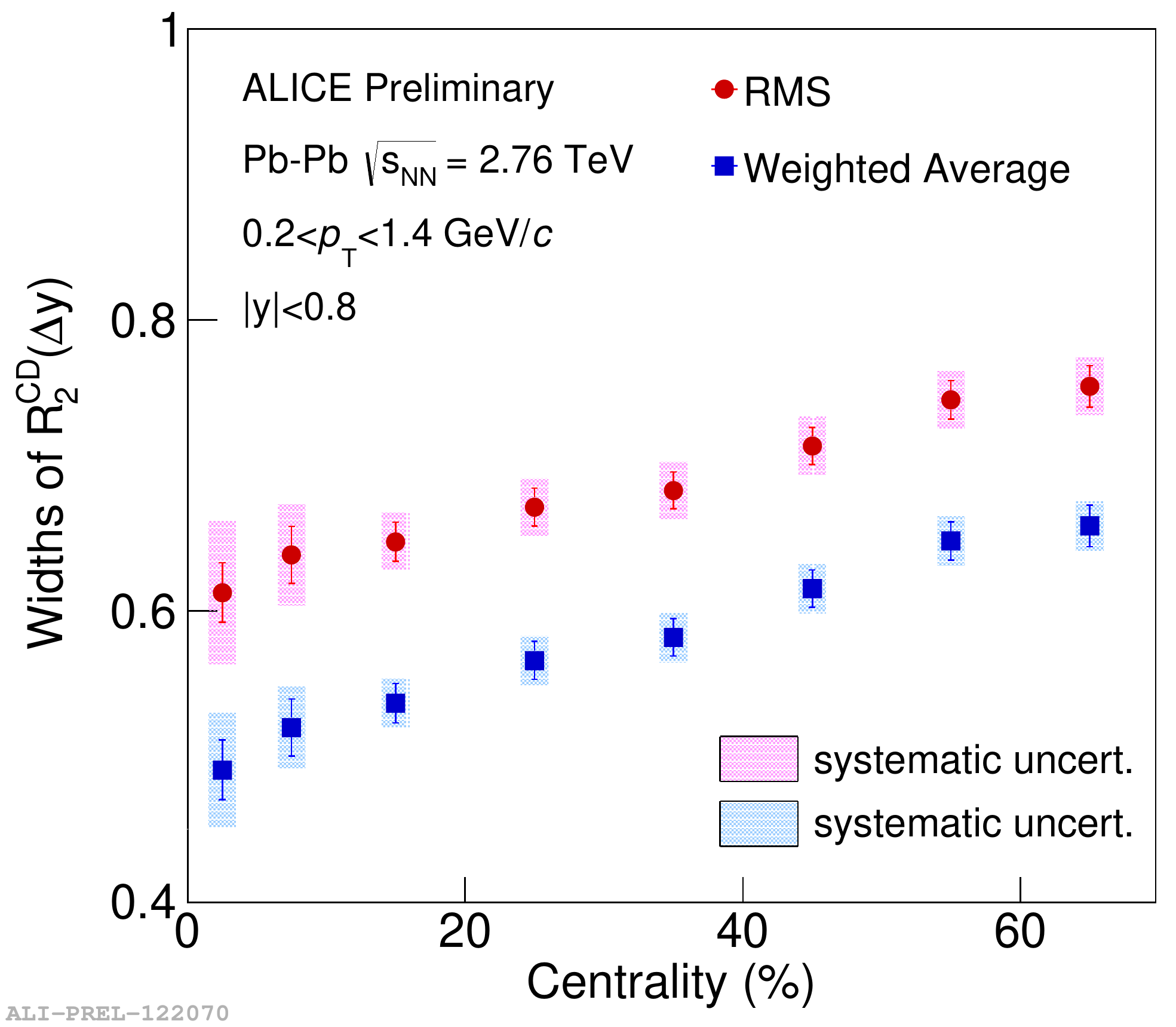} \\
  \end{tabular}
  \caption{The width of the balance function for pions in terms of $\langle \Delta y \rangle $ and $\langle \Delta\varphi \rangle $ as a function of centrality. Error bars shown are systematic. The width in $\Delta\varphi$ is shown for different ranges in $\Delta\varphi$ when projecting.}
  \label{fig:4}

\end{figure}

The Width of the BF provides information on the hadronisation time of particles produced in the collisions. Width is calculated using the RMS value or the weighted average of the BF distributions projected in $\Delta y$ and $\Delta\varphi$. The results for identified charged pion pairs are shown in Fig.\ref{fig:3}. The correlation functions have a dip near $\Delta y$ = 0 and $\Delta\varphi$ = 0. This dip might be a combined effect of Bose-Einstein correlations and Coulomb interactions between charged pions. Widths of the balance functions for three centrality classes are shown in Fig.\ref{fig:4}. A significant narrowing of the BF width of pions with increasing centrality is observed. The broadening of the balance functions for less central collisions is a result of a larger separation between balancing charges. This centrality dependence is consistent with two scenarios: delayed hadronisation and increasing radial flow.

\section{Discussions}
The widths of the balance functions for pions in ∆y and ∆φ are measured in Pb-Pb collisions
at $\sqrt{s_{NN}}$ = 2.76 TeV with the ALICE experiment. They are found to decrease when moving from peripheral to central collisions consistent with the picture of a delayed hadronisation scenario. On the other hand, radial flow should also produce a narrower balance function in central collisions where radial flow is the largest. Therefore, the interpretation of the observed narrowing trend requires more detailed study of its sensitivity to such other effects as flow, resonance production, and diffusion, in addition to late stage hadronisation.

\end{document}